\documentstyle[graphicx,float]{article}

\begin{document}
\title{Two-particle Kapitza-Dirac diffraction}
\author{Pedro Sancho \\ Delegaci\'on de AEMET en Castilla y
Le\'on  \\ Ori\'on 1, 47014, Valladolid, Spain}

\date{}
\maketitle
\begin{abstract}
We extend the study of Kapitza-Dirac diffraction to the case of
two-particle systems. Due to the exchange effects the shape and visibility of
the two-particle detection patterns show important differences for
identical and distinguishable particles. We also identify a novel quantum
statistics effect present in momentum space for some values of the
initial particle momenta, which is associated with different
numbers of photon absorptions compatible with the final momenta.
\end{abstract}


\section{Introduction}

A long time ago Kapitza and Dirac proposed that a standing-wave
light field can act as a diffraction grating \cite{KD}. This
proposal is interesting in two main aspects. On the one hand, from
the fundamental point of view, it provides a nice demonstration of
the wave-particle duality where the diffraction grating is not
massive, but made of massless entities, the photons. On the other
hand, from a more practical perspective, it allows for the design of
matter-wave interferometers (see, for instance, \cite{Ada}). The
effect has been experimentally observed, first with atomic beams
\cite{Gou,Mar}, later with cold atoms \cite{Cah}, and finally with
electrons \cite{Ba1,Ba2}.

More recently, following the seminal work of Hong-Ou-Mandel (HOM)
\cite{HOM}, the effects associated with the quantum statistics of the
particles have been extensively studied in many-particle
interferometry. These studies were mainly concerned with two-photon
interferences originated by the interaction in a beam splitter.
Later, it has been suggested that the exchange effects can also play
an important role in interferometry of identical massive two-particle
systems by diffraction gratings \cite{PS}. In particular, in that paper it
was shown that the diffraction patterns originated at a single slit
are very different for distinguishable particles and fermions and
bosons.

It seems natural to study the behavior of two-particle systems
interacting with the Kapitza-Dirac arrangement. However, this is an
almost unexplored subject. Up to our knowledge, only in \cite{mey} has a
numerical simulation of some basic aspects of the problem for
identical particles been considered. We want to present in this
paper a more general treatment of the subject for two-particle
systems, emphasizing the differences with distinguishable
particles. We shall mainly analyze two aspects of the problem: the
spatial dependence of the detection patterns, and the exchange
effects present in momentum space. With respect to the first point
we shall find that, as in the case of a single slit reported in
\cite{PS}, the patterns show notorious differences for
distinguishable and identical particles, in particular, for the
shape and visibility of the interference figures. A similar
behavior is also found for the correlation functions. The second
aspect, exchange effects in momentum space, is a less studied
subject. We shall show that for some values of the initial momenta
of the particles a novel effect is present in our system. It is
reflected in the changes experienced by the probabilities of finding
the particles in some final states. The physical mechanism
underlying these changes is the possibility of different numbers of
photon absorptions (the mechanism in the basis of the diffraction of
the particles) compatible with the final state of the particles.
This effect occurs in addition to the usual bunching and
antibunching effects in momentum space, which take place for close
values of the momenta.

The calculations will be first carried out for single-mode states to present the main ideas in a simple way. Later, we shall
move to the more realistic case of multi-mode states.

The plan of the paper is as follows. In Sect. 2 we present the basic
equations. The single-mode diffraction patterns are evaluated in
Sect. 3, where we also briefly discuss the behavior of correlation
functions. Section 4 deals with the same problem, but for multi-mode
states. In Sect. 5 we move to momentum space, in which we describe
the existence of novel exchange effects for some values of the
parameters. Finally, in Sect. 6, we discuss the principal results of
the paper.

\section{General expressions}

We consider a pair of particles, distinguishable or not, interacting
with an optical standing wave, usually a laser beam. The wave acts
as a Kapitza-Dirac diffraction grating (see Fig. 1). After the
grating we place detectors measuring the interference pattern, which
can be obtained after many repetitions of the experiment.

The particles passing through the standing-wave light experience a
potential of the form $V=V_0 \cos ^2 k_Lx$, with $k_L$ the
wavenumber of the light field and $x$ the coordinate in the
direction parallel to the light beam \cite{HBa}. The wavefunction
after the interaction can easily be calculated using standard
techniques (see \cite{HBa} for the diffraction and Bragg regimes and
\cite{Fed} for the intermediate region between them). For the short
interaction times between the particles and the light field the
Raman-Nath approximation, equivalent to neglect the particle motion
during the interaction, holds \cite{Ada}. This approximation is
usual in the diffraction regime, the only one we shall consider in this
paper. The single-mode wavefunction of a particle prior to the
interaction is given by $\psi (x,X,t=0) = \exp (i(k_0x+ K_0 X))$,
with $X$ the coordinate of the direction perpendicular to the
optical wave, and $k_0$ and $K_0$ the initial wavenumbers. The
evolved wavefunction after the interaction is given by
\begin{equation}
\psi (x,X,t)= e^{-iVt/\hbar} \psi (x,X,t=0)=e^{iK_0X} e^{-iV_0t \cos ^2 k_L x/\hbar }e^{ik_0x}
\end{equation}
where $t$ is the interaction time. The evolution in the $X$-axis is not modified by the interaction.
\begin{figure}[H]
\center
\includegraphics[width=6cm,height=6cm]{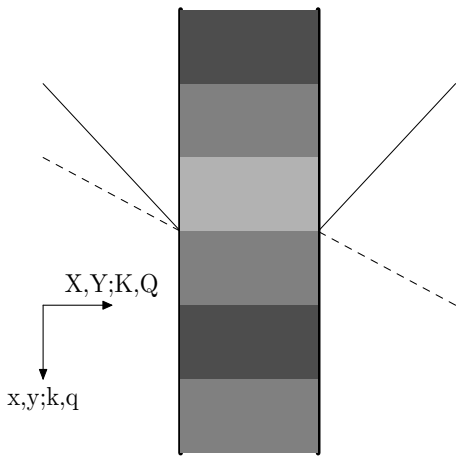}
\caption{Schematic representation of the arrangement. The continuous
and dashed lines represent the two particles}
\end{figure}
Recalling the well-known expressions $\exp (iz \cos \varphi )=\sum
_n i^n J_n(z) \exp (in\varphi )$, with $J_n$ the Bessel functions,
and $\cos ^2 \varphi = \frac{1}{2}+\frac{1}{2} \cos 2\varphi $, we
easily obtain
\begin{equation}
\psi (x,X,t)= e^{iK_0X} \sum _{n=-\infty}^{\infty } b_n e^{i(2nk_L + k_0)x}
\label{eq:dos}
\end{equation}
with $b_n=i^n e^{-iw}J_n(-w)$ and $w =V_0t/2\hbar $.

The wavenumber of the particle can only be modified by double
recoils. For atoms, the first one is associated with the photon
absorption whereas the second one corresponds to stimulated
emission. In the case of electrons, the double scattering can be
understood as a stimulated Compton scattering \cite{HBa}.

We consider now the case of two particles. We denote by $\psi
_{k_0 K_0} (x, X,t)$ and $\psi _{q_0 Q_0} (y,Y,t)$ two
wavefunctions, with an obvious notation. When the particles are
identical the usual product wavefunction $\Psi (x,X,y,Y,t) = \psi
_{k_0 K_0} (x, X, t)\psi _{q_0 Q_0} (y,Y,t)$, valid for
distinguishable particles, must be replaced by
\begin{equation}
\Psi (x,X,y,Y,t) = \frac{1}{\sqrt{2}} \left( \psi _{k_0 K_0} (x, X,t)\psi _{q_0 Q_0} (y,Y,t) \pm  \psi _{k_0 K_0} (y, Y,t)\psi _{q_0 Q_0} (x,X,t) \right)
\end{equation}
In the double sign expressions the upper one holds for bosons and
the lower one for fermions.

From the above expressions one can see that, in the case of identical
Particles, we have simultaneously two different interference effects.
On the one hand, for both distinguishable and identical particles,
the distributions $|\psi _{k_0 K_0} (x, X,t)|^2$ display the
interference effects associated with the diffraction grating. On the
other hand, for identical particles we have another interference
effect that is not present for distinguishable ones. This follows
immediately from the term $\pm 2Re(\psi _{k_0 K_0}^* (x, X,t) \psi
_{q_0 Q_0}^* (y,Y,t) \psi _{k_0 K_0} (y, Y,t) \psi _{q_0 Q_0}
(x,X,t))$, contained in the expression of $|\Psi (x,X,y,Y,t)|^2$.
This form agrees with the standard interpretation of exchange
effects as interference effects \cite{fan}.

\section{Spatial probability distributions}

In this section we evaluate the spatial distribution of simultaneous
two-particle detections and the correlation functions associated
with it. The simplest experimental implementation of the arrangement
consists of two detectors, one fixed at a given position and the
other placed at different points in successive repetitions of the
experiment. In a first step we restrict our calculations to
single-mode states, postponing the discussion of multi-mode ones to the next section.

The evaluation of the probability distribution is simple. We assume
the detection time is fixed at $t$ and we can drop the temporal
variable from all the expressions. We rewrite Eq. (\ref{eq:dos}) as
\begin{equation}
\psi (x,X)= e^{iK_0X}e^{ik_0x} \phi (x)
\end{equation}
with
\begin{equation}
\phi (x)= \sum _{n=-\infty}^{\infty } b_n e^{i2nk_L x}
\end{equation}
With this notation the two-particle probability distribution becomes
\begin{equation}
|\Psi _{dis}(x,X,y,Y)|^2=|\phi (x)|^2|\phi (y)|^2
\end{equation}
for distinguishable particles, and
\begin{equation}
|\Psi (x,X,y,Y)|^2=|\phi (x)|^2|\phi (y)|^2(1 \pm \cos ((K_0-Q_0)(X-Y)+(k_0-q_0)(x-y)))
\end{equation}
for identical ones.

We assume from now on that $X=Y$, i. e., the positions of the detectors
in the direction perpendicular to the light field to be equal. Several
consequences easily emerge from the above expressions. In the case
of identical particles, the distribution is the product of the {\it
distinguishable distribution} by the term $1 \pm \cos
((k_0- q_0)(x-y))$. The first one reflects the two-particle
interferences associated with the dispersion by the light field. It
is a function, through $\phi$, of $k_L$ and $b_n(V_0t)$ the
parameters of the optical diffraction grating. That dependence is
similar for both identical and distinguishable particles. On the other
hand, the term containing the cosine function is related to the exchange
effects. For fixed $x$ and $y$ it is only function of the initial
momenta of the particles in the direction of the light field. The
bunching and antibunching effects directly emerge from the above
equations. In the case of bosons, for $x \approx y$ we have $|\Psi
_B(x,X,y,X)|^2 \approx 2|\phi (x)|^2|\phi (y)|^2$, i. e., the
probability of two-boson detection almost doubles that of two
distinguishable particles. For double detection at the same point,
$x=y$, we have, as discussed in \cite{PS}, $|\Psi _B(x,X,x,X)|^2 =
2|\phi (x)|^4$, i. e., a dependence on the fourth power of the
wavefunction modulus. By contrast, for fermions we have for $x
\approx y$, $|\Psi _F(x,X,y,X)|^2 \approx 0$ that is the
antibunching effect. When $x=y$ the probability is strictly null
according to the exclusion principle. There is another situation in
which the two-fermion detection probability is, for
any $x$ and $y$, identically null. It occurs for equal initial momenta in
the direction of the light field, $q_0=k_0$. This is so even when
$K_0 \neq Q_0$ (in the case of $K_0 = Q_0$ it is evident that the
distribution must be null because it is impossible to prepare two
fermions in the same state).

Next we present a graphical representation of the above equation. As
discussed before we fix one of the detectors at $y=0$ and move the
other at different positions $x$. A simple calculation gives the
explicit form for $\phi (x)$ and its squared modulus:
\begin{equation}
|\phi (x)|^2=1+\sum _{n,m}^{m>n} 2(-1)^{n+m} J_n(w) J_m(w ) \cos \left( (m-n) \left( 2k_Lx + \frac{\pi}{2} \right) \right)
\end{equation}

In the calculation we have used the properties $\sum _n |b_n|^2=1$
that derives directly from the normalization condition, and
$J_n(-w)=(-1)^n J_n(w)$.

The normalization of $\Psi _{dis}$ is automatically guaranteed by
the normalization of $\phi $. On the other hand, for identical
particles $|\Psi |^2$ contains terms of the form $\cos ((m-n)2k_Lx +
\pi /2) \cos ((k_0 - q_0)x)$. The integration over $x$ of these
terms is not null (see, for instance, \cite{PS} where similar
integrations are carried out) when $(m-n)2k_L=\pm (k_0 - q_0)$. When
this condition is fulfilled the normalization is not given by $|\phi
(0)|^2\sum _n |b_n|^2=1$, but by the expression $|\phi (0)|^2(\sum
_n |b_n|^2 \pm I)=1$ with $I$ the result of the above integration.
\begin{figure}[H]
\center
\includegraphics[width=6cm,height=6cm]{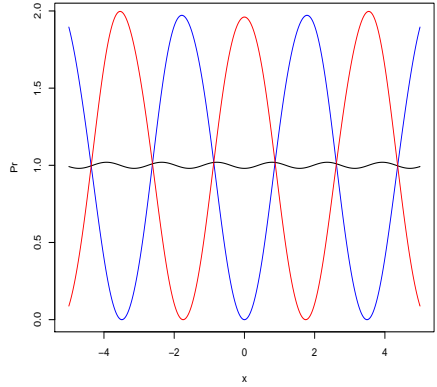}
\caption{In the vertical axis we represent the two-particle
detection probability and in the horizontal one the coordinate $x$.
The red, black and blue curves correspond, respectively, to bosons,
distinguishable particles and fermions. We use the values $w =0.2$,
$k_0=0.9$, $q_0=-0.9$, and $k_L=1$.}
\end{figure}
The figure shows a large increase of the visibility of the
two-particle detection probabilities for identical particles. As usual the visibility is defined as
\begin{equation}
{\cal V} = \frac{|\Psi (x,X,0,X)|_{max}^2 -|\Psi (x,X,0,X)|_{min}^2}{|\Psi (x,X,0,X)|_{max}^2 + |\Psi (x,X,0,X)|_{min}^2}
\end{equation}
In the case of distinguishable particles the probability
approximately oscillates between $0.98$ and $1.02$, whereas for
identical ones it does between $0$ and $2$. Then we have,
respectively, ${\cal V}_{dis} \approx 0.02$ and ${\cal V}_{ide}
\approx 1$, i. e., a very large difference.

The curves for bosons and fermions are very similar, with small
changes of intensity due to the modulation introduced by the
multiplicative factor $|\phi (x)|^2|\phi (0)|^2$ and a phase-$\pi $
displacement associated with the sign $\pm $ (which explains the
difference between bunching and antibunching at $x \approx 0$).

Now, we briefly analyze the behavior of the correlation functions.
These functions have been extensively used in the experimental study
of the (anti)bunching effect of free particles released
from optical lattices or magnetic traps \cite{Jel,Fol,Rom}. In
particular, in the last two references the existence of periodic
correlations was shown between density fluctuations of atoms
released from optical lattices. These correlations reflect the
underlying structure of the lattice.

The correlation function is defined as
\begin{equation}
C(\eta )=\frac{ 1}{d}\int _{0}^{d} |\Psi (x,X,x+\eta ,X)|^2 dx
\end{equation}
with $d=2\pi /k_L$ representing the spatial periodicity of the
optical diffraction grating. In our case, this expression transforms
into
\begin{equation}
C(\eta )=\frac{1}{d}(1 \pm \cos ((q_0-k_0)\eta )) \int _{0}^{d}
|\phi (x)|^2|\phi (x + \eta )|^2 dx
\end{equation}
The evaluation of the integral is simple but lengthy. We only
present the final result (a similar calculation can be found in
\cite{PS}):
\begin{eqnarray}
C(\eta )=(1 \pm \cos ((q_0-k_0)\eta )) \times \nonumber \\
(1+ 2 \sum _{\cal N} (-1)^{n+m+r+s} J_n(w) J_m(w) J_r(w) J_s(w) \cos (2(r-s)k_L \eta ))
\end{eqnarray}
with ${\cal N}$ denoting the set of $n,m,r,s$ with the constrains $m>n$, $s>r$ and $m-n = \pm (r-s)$.

This correlation function is the product of two periodic functions.
The first one, that in the left side, is only relevant for identical
particles. For distinguishable ones it reduces to $1$ and,
consequently does not vary for different initial preparations of the
pairs. For identical particles it only depends on their initial
momenta. For bosons it reaches maximum values for any $\eta = 2N\pi
/(q_0-k_0)$ with $N$ an integer and minimum ones for $\eta = (2N+1)
\pi /(q_0-k_0)$. A similar behavior holds for fermions when an
additional phase-$\pi$ is taken into account. On the other hand, the
second multiplicative factor is similar for both distinguishable and
identical particles and bosons and fermions. It only depends on the
parameters of the optical grating, $w$ and $k_L$. As all the terms
of the type $(r-s)k_L$ are contained in this factor, we have in the correlation
function all the periods generated by the optical grating. As
discussed in \cite{Fol,Rom,PS}, it reflects the underlying structure
of the diffraction grating.

\section{Multi-mode states}

Up to now, we have only considered single-mode states. Now, we move
to the more realistic case of multi-mode ones. The simplest way to
study them is to assume that the distribution of initial wavenumbers
of each particle is a Gaussian one. The one-dimensional Gaussian
distribution is given by $f(k_0)=(4\pi )^{1/4} \sigma ^{-1/2} \exp
(-(k_0-\Lambda )^2/2 \sigma ^2)$ with $\sigma $ the width of the
distribution and $\Lambda $ its central value. The initial
wavefunction reads as $\psi (x,t=0)=\int dk_0 f(k_0)e^{ik_0 x}$.
Note that, for the sake of simplicity, we have not included the
variable $X$, which is irrelevant in the following discussion. The
wavefunction after passing through the light grating is:
\begin{eqnarray}
\psi (x)=\sum _{n=-\infty }^{\infty } b_n e^{i2nk_L x} \int dk_0
f(k_0) e^{ik_0 x} \nonumber \\
\sim e^{-x^2 \sigma ^2 /2} \sum
_{n=-\infty }^{\infty } b_n e^{i2(nk_L +\Lambda ) x} = e^{-x^2
\sigma ^2 /2} \psi _{\Lambda } (x)
\end{eqnarray}
where we have carried out a trivial integration over $k_0$, and $\psi
_{\Lambda }$ represents the wavefunction of a particle with initial
wavenumber $\Lambda $. For the matter of simplicity we have not
included the constant coefficients of the distribution and those
derived from the integration. We conclude that the multi-mode
wavefunction after the interaction equals that of a single-mode
particle with the central value of the distribution, but spatially modulated by a Gaussian distribution.

In the next step we consider two particles in multi-mode states,
with central values $\Lambda $ and $\Upsilon$ and widths $\sigma $
and $\mu $. The probability distribution after the interaction is
\begin{eqnarray}
|\Psi (x,y)|^2 = \frac{1}{2}e^{-x^2 \sigma ^2 } e^{-y^2 \mu ^2}|\phi (x)|^2 |\phi (y)|^2 + \nonumber \\
\frac{1}{2} e^{-y^2 \sigma ^2 } e^{-x^2 \mu ^2 }|\phi (y)|^2 |\phi (x)|^2  \pm \nonumber  \\
e^{-(x^2+y^2)( \sigma ^2 +\mu ^2)/2} |\phi (x)|^2 |\phi (y)|^2 \cos((x-y)(\Lambda - \Upsilon))
\end{eqnarray}
We represent this distribution. We take the same values of Fig. 2,
in particular, $\Lambda =k_0$ and $\Upsilon =q_0$. For the width of the distribution we take $\sigma ^2 =\mu ^2 = 0.2$.
\begin{figure}[H]
\center
\includegraphics[width=6cm,height=6cm]{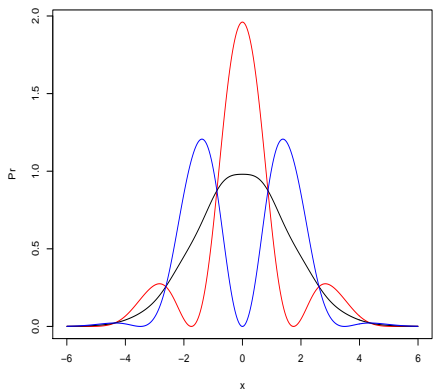}
\caption{The same as in Fig. 2, but for multi-mode states.}
\end{figure}
As in the single-mode case, there is a notorious difference between
the curves of distinguishable and identical particles. For the first
one, there is an almost flat distribution in the center followed by
an exponential-like decreasing superposed with small oscillations
(associated with $|\phi (0)|^2|\phi (x)|^2$). On the other hand, for
identical particles there is an interference pattern. In the
proximities of the point $x=0$ we observe the bunching and
antibunching effects. The values of the visibility are clearly
different for the three figures, but not so much as in the
single-mode case. These contrasts in the visibility are strongly
enhanced for small values of $\sigma $.

\section{Momentum space}

In the two previous sections we have studied the spatial
two-particle interference patterns. Now we consider the momentum
space to see how the exchange effects manifest in this
representation. The methodology is similar to that used above, first
we consider single-mode states where the principal ideas can be
analyzed in a simple way, and later we address the multi-mode case.

The wavefunction in momentum space (the momentum and wavevnumber are
related by the trivial relation $p=\hbar k$, and it is justified to
speak about the momentum space although really we are dealing with
wavevenumbers) is given by the Fourier transform of the wavefunction in the
position representation, $\Phi (k)=\int dx e^{-ikx} \psi (x)$.
For the identical two-particle system we have
\begin{equation}
\Phi (k,q)=\int dx \int dy e^{-i(kx+qy)} \Psi (x,y) = \frac{1}{\sqrt 2} \Phi _{k_0} (k) \Phi _{q_0} (q) \pm \frac{1}{\sqrt 2} \Phi _{k_0} (q) \Phi _{q_0} (k)
\end{equation}
where the single-particle wavefunctions are characterized by the
value of the initial wavenumber (for multi-mode states by the
central value of the distribution).

The experimental variable to measure is the probability of detecting
one particle with the value $k$ and the other with $q$. In the
arrangement we must replace the detectors of the previous sections
by momentum measurement devices. In a large series of
measurements, for instance fixing $k$ and scanning for different
values of $q$, we can compare the experimental data with the
theoretical distribution:
\begin{eqnarray}
|\Phi (k,q)|^2=\frac{1}{2} |\Phi _{k_0} (k)|^2 |\Phi _{q_0} (q)|^2 + \frac{1}{2} |\Phi _{k_0} (q)|^2| \Phi _{q_0} (k)|^2 \nonumber \\
 \pm  Re (\Phi _{k_0}^* (k) \Phi _{q_0}^* (q) \Phi _{k_0} (q) \Phi _{q_0} (k))
\end{eqnarray}
This expression shows that the probability distributions in momentum
space have a structure similar to that in the position
representation. In particular, for $k \approx q$ we have the
bunching and antibunching effects in momentum space. This effect has
been numerically discussed in \cite{mey}.

To get a better understanding of that distribution we
shall consider specific examples, starting with the single-mode states.

\subsection{Single-mode states}

We consider successively the cases of a single particle, two distinguishable
Particles, and two identical ones.

{\bf One particle.} This case is particularly simple because the
wavefunction reduces to a superposition of Dirac's deltas:
\begin{equation}
\Phi (k)= \sum _n b_n \delta (k-2nk_L - k_0)
\end{equation}
For a single particle the probability of finding in a momentum
measurement the value $2nk_L + k_0$ is $|b_n|^2$ because, in the
sense of the distributions, for $r \neq n$ the terms $\delta
(k-2nk_L - k_0) \delta (k-2rk_L - k_0)$ are zero.

{\bf Two distinguishable particles.} Now, the probability distribution is
\begin{equation}
|\Phi _{k_0}(k)|^2|\Phi _{q_0}(q)|^2= \sum _{n,m} |b_n b_m|^2 \delta (k-2nk_L - k_0) \delta (q-2mk_L - q_0)
\end{equation}

The probability of obtaining in simultaneous measurements the values
$2nk_L + k_0$ and $2mk_L + q_0$, denoted as $P(n,m)$, is $|b_n
b_m|^2$.

To illustrate the form of this probability we graphically
Represent it in Fig. 4 it for the lower values of $n$ and $m$. Note that this
probability is equal to the probability of $n$ absorptions by the
particle with initial wavenumber $k_0$ and $m$ by that with $q_0$.
For small values of $w$ there is only one absorption. When $w$
increases the probability of two absorptions becomes dominant. The
probability of symmetric absorption (one photon each particle,
$P(1,1)$) is much larger than the asymmetric one (one particle two
photons and the other none, $P(0,2)$). For larger values of $w$ the
terms with three absorptions become important. However, here, there
is again a notorious difference between $P(1,2)$ that increases with
$w$ and $P(0,3)$ that remains negligible for all the range of values
considered. At the end of the graphic, the term $P(2,2)$ becomes
comparable to (or larger than) the other terms.

\begin{figure}[H]
\center
\includegraphics[width=6cm,height=6cm]{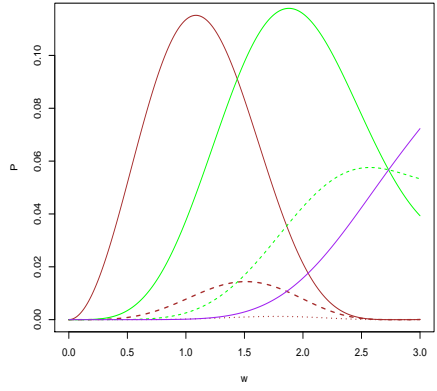}
\caption{Representation of the probability P(n,m) versus w. The
brown curves correspond to $P(0,1)$ (continuous), $P(0,2)$ (dashed),
and $P(0,3)$ (dotted), the green ones to $P(1,1)$ (continuous), and
$P(1,2)$ (dashed), and the purple one to $P(2,2)$.}
\end{figure}

{\bf Two identical particles.} The two direct terms in $|\Phi
(k,q)|^2$ are similar to those for distinguishable particles. We
evaluate now the term associated with the exchange effects:
\begin{eqnarray}
\Phi _{k_0}^* (k) \Phi _{q_0}^* (q) \Phi _{k_0} (q) \Phi _{q_0} (k) = \nonumber \\
\sum _{n,m,r,s} b_n^* b_m^* b_r b_s  \delta (k-2nk_L - k_0) \times \nonumber \\ \delta (q-2mk_L - q_0) \delta (q-2rk_L - k_0) \delta (k-2sk_L - q_0)
\end{eqnarray}
The product of the two deltas containing $k$ is zero (in the sense
of the distributions) unless the relation $2nk_L + k_0 =2sk_L + q_0$
holds. Similarly, we must have $2mk_L + q_0 =2rk_L + k_0$ in order for
the product of the deltas containing $q$ not be null.
These two relations can be expressed as
\begin{equation}
n-s=N  \; ; \; m-r=-N \; ; \; N=\frac{q_0 - k_0}{2k_L}
\end{equation}
with $N$ an integer. The difference between the initial wavenumbers
must be an integer number of times $2k_L$.

When these relations hold, we have that the terms related to the deltas containing $n$ and $m$ must be rewritten as
\begin{eqnarray}
\frac{1}{2}|b_n b_m|^2 \delta (k-2nk_L - k_0) \delta (q-2mk_L - q_0)+ \nonumber \\
\frac{1}{2}|b_n b_m|^2 \delta (q-2nk_L - k_0) \delta (k-2mk_L - q_0) \pm \nonumber \\
Re(b_n^*b_m^*b_{m+N} b_{n-N}) \delta (k-2nk_L - k_0) \delta (q-2mk_L - q_0)
\end{eqnarray}
The two first terms correspond to the absorption of $n$ photons by a
particle with initial wavenumber $k_0$ and $m$ by one with $q_0$. If
$ (q_0 - k_0)/2k_L$ is not an integer, the third term does not
contribute and we have the two final wavenumbers $2nk_L+k_0$ and
$2mk_L+q_0$ with probability $|b_nb_m|^2$. On the other hand, when
the condition holds we obtain the same final wavevenumbers
$2nk_L+k_0$ and $2mk_L+q_0$, but with probability
\begin{equation}
P_N (n,m) = |b_n b_m|^2 \pm Re(b_n^*b_m^*b_{m+N} b_{n-N})
\end{equation}
This result can easily be understood in terms of
indistinguishability of alternatives. If the final wavenumbers are
the same for both types of absorptions, we cannot distinguish if the
final result corresponds to the alternative of absorptions $n$ and
$m$, or to the other alternative, $n-N$ and $m+N$. In quantum theory
the amplitudes of probability for indistinguishable alternatives
must be added, obtaining an interference effect. The
indistinguishable alternatives correspond to different numbers of
absorptions yielding the same final wavenumbers. The situation is
different for distinguishable particles. In this case, we can know
in principle at any instant if the particle is of one type (that
whose wavenumber is denoted by $k$) or the other (wavenumber $q$)
and, consequently, if its initial wavenumber was $k_0$ or $q_0$.
Then one particle of the first kind cannot reach the final
wavenumber $2nk_L+k_0$ by $n-N$ absorptions. There are not different
alternatives to reach the final wavenumber and, there is not an
interference effect.
\begin{figure}[H]
\center
\includegraphics[width=10cm,height=10cm]{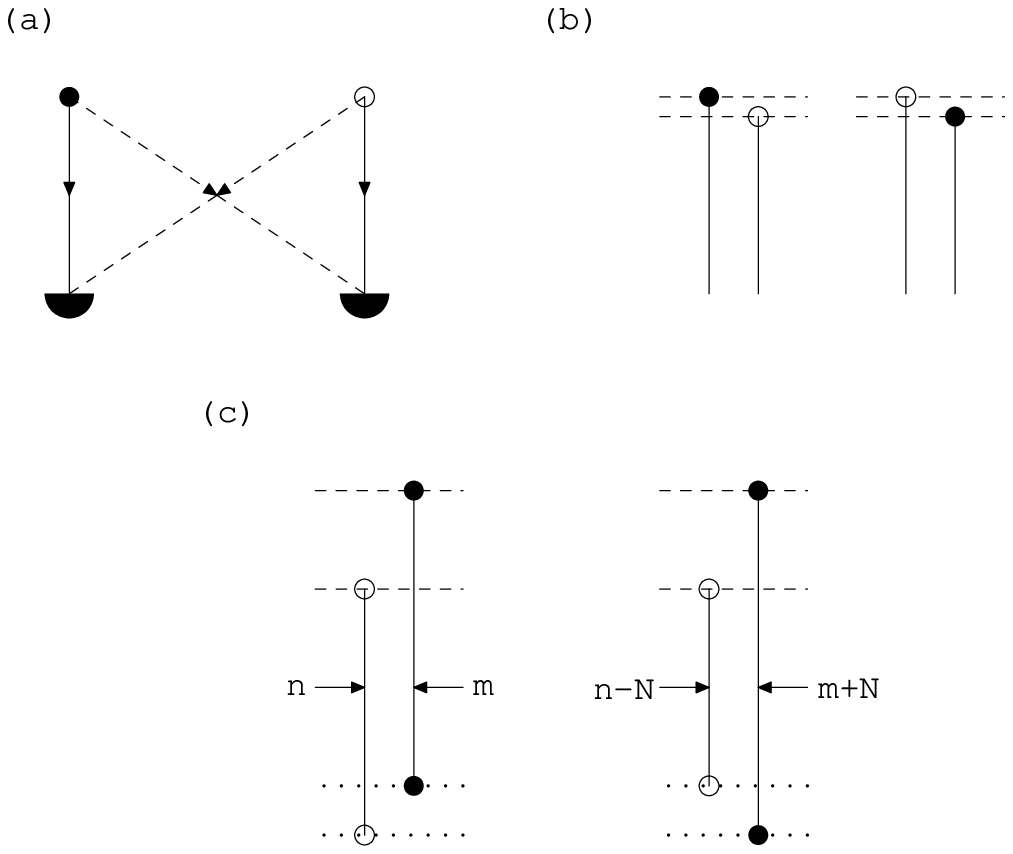}
\caption{Representation of the exchange effects. Black and white
circles represent particles labeled by $k$ and $q$. (a) The standard
(anti)bunching effect in the spatial representation. The detections
can occur in two alternative ways, these represented by the continuous
and dashed lines. (b) The same effect in momentum space. The dashed
horizontal lines correspond to the two momenta at which the
particles are detected and the continuous vertical ones to the
trajectories in momentum space. The (anti)bunching is observed when
the two final momenta are very close. (c) The effect discussed in
this paper. The dotted lines correspond to the initial momenta. The
arrows represent the absorption of photons. Being the particles
identical the two represented alternatives are indistinguishable.}
\end{figure}
We remark that this novel exchange effect is different from
(anti)bunching (see Fig. 5). The last one only takes place when we
consider particles very close in momentum space. In contrast, the
effect described here only depends on the initial values of the
momenta (which can be rather different for all the cases with $N
\neq 0$) and can occur for well-separated final momenta.

The previously described effect is a purely quantum one. To justify
this statement we show that it cannot be described by a classical
treatment. We use a qualitative model where the particles collide
with the photons and the collisions are ruled by the classical law
of momentum conservation (classical description of the Compton
effect assuming the existence of photons). We assume that all of the
momentum of the photon is transferred to the particle, in such a way
that its momentum changes from $p_0$ to $p_0 + \hbar k_L$. In the
cases with $2n$ collisions the final momentum of the particle would
be the same predicted by quantum theory. However, in the absence of
additional unnatural assumptions, the classical theory allows for
odd numbers of collisions. The agreement between the model and
quantum theory would be even worse for couples of particles.
Classically, both particles behave in an independent way, and the
stochastic collision events are uncorrelated. The probability of one
particle experiencing $2n$ collisions and the other $2m$ factorizes.
The result would be similar to that for quantum distinguishable
particles. In conclusion, in the classical model (a reasonable one)
there is not room for the previous exchange effect.

Next we represent these probabilities for a particular example. A
simple calculation using the expression for $b_n$ and the property
of the Bessel functions $J_{-n}(w)=(-1)^n J_n (w)$ gives $P_0 (1,0)=
J_1^2 J_0^2 \pm J_1^2 J_0^2$, $P_1 (1,0)= J_1^2 J_0^2 \pm J_1^2
J_0^2$ and $P_{-1} (1,0)= J_1^2 J_0^2 \mp J_1^2 J_0 J_2$. We see
that $P_0 (1,0)=P_1 (1,0)$.
\begin{figure}[H]
\center
\includegraphics[width=6cm,height=6cm]{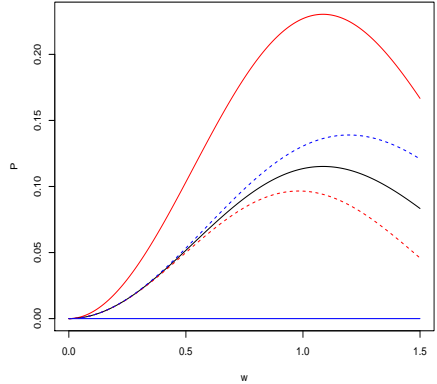}
\caption{Representation of the probability P of finding the final
wavenumbers $2nk_L + k_0$ and $2mk_L + q_0$ versus w. The black
curve represents the distinguishable case for $n=1$ and $m=0$. The
red and blue ones correspond to bosons and fermions with $N=1$
(continuous) and $N=-1$ (dashed).}
\end{figure}

The figure clearly shows that the probabilities of finding the
particle in the targeted final wavenumber state are different for
the all the cases. For $N=1$ the detection rate notoriously
increases for bosons, whereas it becomes null for fermions. The
detection rate is not always enhanced for bosons; for $N=-1$ it is
smaller than that of distinguishable particles. Similarly, the
fermion rate can increase with respect to that of distinguishable
particles (also $N=-1$).

\subsection{Multi-mode states}

In this subsection we analyze how the above results are modified in
the more realistic case of multi-mode distributions. We mainly
restrict our considerations, as in Sect. 4, to Gaussian
distributions which can be tackled analytically. As in the previous
subsection we consider successively the cases of  a single particle,
two distinguishable particles, and two identical ones.

{\bf One particle.} The wavefunction in momentum space for a
particle with a multi-mode distribution $f(k_0)$ of initial
wavenumbers is easily obtained:
\begin{eqnarray}
\Phi (k) = \int dx e^{-ikx} \psi (x)= \int dx e^{-ikx} \sum _n \int dk_0 f(k_0) b_n e^{i(2nk_L+k_0)x} = \nonumber \\
\sum _n \int dk_0 f(k_0) b_n \delta (k-2nk_L-k_0)=\sum _n b_n f(k-2nk_L)
\end{eqnarray}
The probability of finding the particle with final wavenumber $k$ is
$|\Phi (k)|^2= \sum _{n,m} b_n^* b_m f^*(k-2nk_L)f(k-2mk_L)$. If the
mode distribution is a Dirac's delta (the single-mode distribution),
$f(k_0)=\delta (k_0 - K_0)$, the probability becomes $|\Phi (k)|^2=
\sum _n |b_n|^2 \delta(k-2nk_L-K_0)$.

When the initial distribution is a Gaussian, $f(k_0) \sim e^{-(k_0-\Lambda )^2/2\sigma ^2}$, we have
\begin{equation}
|\Phi (k)|^2 \sim \sum _{n,m} b_n^* b_m e^{-(k-2nk_L-\Lambda )^2/2\sigma ^2} e^{-(k-2mk_L-\Lambda )^2/2\sigma ^2}
\end{equation}
These two exponentials correspond to two Gaussians with the same
width of the initial ones, but centered around $2nk_L +\Lambda $ and
$2mk_L + \Lambda $. Two different regimes can be obtained. For $2k_L
\gg \sigma $ we have that the overlapping between the Gaussians
centered around $2nk_L + \Lambda$ and $2(n \pm 1)k_L + \Lambda$
(and, of course, any $m$ with $|n \pm m|>1$) is negligible. The sum
in the above expression reduces to the diagonal terms, $|\Phi (k)|^2
\simeq \sum _n |b_n|^2 e^{-(k-2nk_L-\Lambda )^2/\sigma ^2}$. After
the interaction, we can only detect the particle with wavenumbers
contained in the Gaussian distributions centered in the points
$2nk_L + \Lambda $ (with $1/\sqrt 2$ times the width of the initial
one). There is not a contribution of the crossed terms to the
probability of detection. In the limit of very peaked Gaussians we
recover the behavior described by Dirac's delta distributions. The
other regime takes place when $\sigma \geq 2k_L$. In this case,
there is a non-negligible overlapping between the distributions
centered around $2nk_L + \Lambda$ and $2(n \pm 1)k_L + \Lambda$
(and, perhaps, the other $m$ with $|n - m|>1$). The crossed terms can no
longer be neglected, leading to interference terms. These
interference terms can be understood by the impossibility of
distinguishing if a particle detected with wavenumber $k$ belongs to
one or the other of the distributions.

{\bf Two distinguishable particles.} The discussion closely follows
that of one particle. If the initial distributions of the particles
are $f(k_0)$ and $g(k_0)$ the probability of finding one particle
with $k$ and the other with $q$ is
\begin{equation}
|\Phi _f (k)|^2 |\Phi _g (q)|^2 \sim \sum _{n,m,r,s} b_n^* b_m b_r^* b_s
f^*(k-2nk_L) f(k-2mk_L)g^*(q-2rk_L)g(q-2sk_L)
\end{equation}
We assume the two distributions to be Gaussian ones. For $2k_L \gg
\sigma $ and $2k_L \gg \mu $ we have that the overlappings between
the Gaussians centered around $2nk_L + \Lambda$ and $2(n \pm 1)k_L +
\Lambda$ and $2nk_L + \Upsilon$ and $2(n \pm 1)k_L + \Upsilon$ are
negligible, becoming the probability $|\Phi _f(k)|^2 |\Phi _g(q)|^2
\simeq \sum _{n,m} |b_nb_m|^2 e^{-(k-2nk_L-\Lambda )^2/\sigma ^2}
e^{-(q-2nk_L-\Upsilon )^2/\mu ^2}$. The probability is the product
of the one-particle distributions without crossed terms. When the
conditions $\sigma \geq 2k_L$ and/or $\mu \geq 2k_L$ hold we have crossed  terms leading to interference effects.

{\bf Two identical particles.} The two-particle probability contains
the terms of the form $|\Phi _f (k)|^2 |\Phi _g (q)|^2$, which can
be treated as before, and the exchange term that transforms into
$\sum _{n,m,r,s}Re(b_n^* f^*(k-2nk_L) b_m^* g^*(q-2mk_L) b_r
f(q-2rk_L) b_s g(k-2sk_L))$. We again consider Gaussian
distributions, for which the exchange term reads proportional to
\begin{eqnarray}
\sum _{n,m,r,s} Re(b_n^* b_m^* b_r b_s) e^{-(k-2nk_L-\Lambda )^2/2\sigma ^2} \times \nonumber \\
e^{-(q-2mk_L-\Upsilon )^2/2\mu ^2} e^{-(q-2rk_L-\Lambda )^2/2\sigma ^2} e^{-(k-2sk_L-\Upsilon )^2/2\mu ^2}
\end{eqnarray}
For the sake of simplicity we take $\sigma = \mu$. For $|2(n-s)k_L
+\Lambda - \Upsilon )| \gg \sigma$ or $|2(m-r)k_L - \Lambda +
\Upsilon )| \gg \sigma$ (for any $n$ and $s$ and $m$ and $r$) the
overlapping between the curves is negligible and the product of the
two distributions with the same argument ($k$ or $q$) is almost
null. The contribution of the exchange terms can be neglected. In
contrast, when $|2(n-s)k_L +\Lambda - \Upsilon )| \leq \sigma$ for
some $n$ and $s$ and $|2(m-r)k_L - \Lambda + \Upsilon )| \leq
\sigma$ for some $m$ and $r$ the product of the distributions cannot
be neglected. The exchange effects become important in this case.
However, the conditions for the presence of the exchange effect are
much less stringent than in the case of single-mode states.

\section{Discussion}

We have analyzed in this work the extension of the Kapitza-Dirac
effect to the case of two-particle systems. The spatial two-particle
detection patterns display notorious differences for distinguishable
particles and fermions and bosons, in particular for the shape and
visibility of the interference figure.

We have also shown the existence of a novel exchange effect in
momentum space. The effect, which modifies the distribution of
particles detected with given momenta, only occurs for some values
of the initial momenta of the particles (or in the case of
multi-mode states for some values of the parameters of the initial
momenta distributions). For multi-mode states these conditions are
much less stringent than for single-mode ones. The verification of
this effect would be interesting in several aspects. Exchanges
effects are a striking manifestation of the departure between
classical and quantum descriptions of physical systems. Any new
example of these differences is worth investigating. The novel
effect has no relation with other exchange effects such as
(anti)bunching, exclusion in atoms, or degeneracy in gases. The
physical underlying mechanisms are different, in our case, different
numbers of photon absorptions. Moreover, the confirmation of the
effect would corroborate the validity of the (anti)symmetrization
principle in a framework where strong interactions with other types
of particles are present.

In this paper, we have only considered the spatial part of the
wavefunction. This is equivalent to assume that the spin states (and
the electronic states for atoms) are symmetric and do not play any
role in the problem. The extension to the case of antisymmetric spin
or electronic states, where the relative sign of the spatial part of
the wavefunction can be reversed, is simple.

Some other aspects of our approach must be commented on. The
wavefunctions have just been evaluated after the interaction with
the optical grating. The detectors must be placed at these
positions, and consequently, we are in the near-field regime.
However, in this type of problems one usually works in the far-field
one \cite{Ba2}. In the free evolution of the particles between both
regimes there is a dispersion of the wavefunctions. Fortunately,
that evolution is simple (free evolution) and can easily be
described. For instance, in the spatial picture we have that the
width of the Gaussian packet increases with time. As the momentum
peaks are determined via spatial detection \cite{Ba2} we must take
into account that broadening to correctly interpret the data. For
well-separated peaks the effects of the spreading are negligible.
For peaks with appreciable overlapping the range of values of the
initial momenta for which the exchange effect described in this
paper takes place increases (as we can easily see with an argument
similar to that used for multi-mode states in subsection 5.2). An
exact treatment of the problem would require a quantitative
evaluation of the evolution of the multi-mode states, instead of the
qualitative one presented here. However, this will be presented
elsewhere. Another important aspect of the problem is the question
of the coherence between the two particles. As is well-known, when
they are only partially coherent the two-particle interference
properties, in particular the visibility, are modified. Thus, the
relative coherence of the two particles should be checked for every
type of source of pairs of particles. When it is only partial we
should introduce the necessary modifications in the formalism. In
our proposal, there is an additional question similar to that of the
partial coherence: we must analyze if the overlapping between the
wavefunctions is complete or partial (see also the next paragraph).
In the first case we can use a completely (anti)symmetrized
wavefunction. In the second one, we should include in the
wavefunction (or in a density matrix) the property of partial
overlapping. This is a subject worth analyzing since, as far as we
know, it is not considered in the literature. Here, by a matter of
simplicity, we have assumed complete {\it coherence} between both
particles in the two senses, coherence and overlapping.

The last question we briefly address here is the possibility of
experimentally testing the above effects. In the space
representation, the interchange effects are present when there is a
non-negligible overlapping between the wavefunctions of the two
particles. Then we must prepare the particles in such a way that
they have a non-negligible overlapping at the time they reach the
optical grating. Several sources of identical particles, such as
Bose-Einstein condensates, optical lattices, or magnetic traps have
recently been used to study correlation functions
\cite{Jel,Fol,Rom}. If we would be able to select the cases in which
only two particles are released from any of the above devices we
would have an efficient source for our problem. We would also
carefully test that the times of the arrival of the two particles to the
optical grating are close enough. In terms of wavepacket spread, the
peaks of the two probability distributions must reach the optical
grating at the same time. On the other hand, the interaction
strength is given, as in the one-particle case, by the parameter
$w$. Using the same values of the single particle arrangement
\cite{HBa} we can reach an appreciable value for the number of pairs
diffracted.

{\bf Acknowledgments} We acknowledge partial support from MEC (CGL 2007-60797).


\begin{thebibliography}{99}
\bibitem{KD} P. L. Kapitza and P. A. M. Dirac, Proc. Cambridge Philos. Soc. {\bf 29}, 297 (1933).
\bibitem{Ada} C. S. Adams, M. Sigel, and J. Mlynek, Phys. Rep. {\bf 240}, 143 (1994).
\bibitem{Gou} P. L. Gould, G. A. Ruff, and D. E. Pritchard, Phys. Rev. Lett. {\bf 56}, 827 (1986).
\bibitem{Mar} P. J. Martin, B. G. Oldaker, A. H. Miklich, and D. E. Pritchard, Phys. Rev. Lett. {\bf 60}, 515 (1988).
\bibitem{Cah} S. B. Cahn, A. Kumarakrishnan, U. Shim, T. Sleator, P. R. Berman,  and B. Dubetsky, Phys. Rev. Lett. {\bf 79} 784 (1997).
\bibitem{Ba1} D. L. Freimund, K. Aflatooni, and H. Batelaan, Nature {\bf 413}, 142 (2001).
\bibitem{Ba2} D. L. Freimund and H. Batelaan, Phys. Rev. Lett. {\bf 89} 283602 (2002).
\bibitem{HOM} C. K. Hong, Z. Y. Ou, and L. Mandel, Phys. Rev. Lett. {\bf 59},  2044 (1987).
\bibitem{PS} P. Sancho, J. Phys. B {\bf 43} 065504 (2010).
\bibitem{mey} G. Lenz, P. Pax, and P. Meystre, Phys. Rev. A {\bf 48}, 1707 (1993).
\bibitem{HBa} H. Batelaan, Contemp. Phys. {\bf 41}, 369 (2000).
\bibitem{Fed} M. V. Fedorov, JETP {\bf 52}, 1434 (1967).
\bibitem{fan} U. Fano, Am. J. Phys. {\bf 29}, 539 (1961).
\bibitem{Jel} T. Jeltes, et al., Nature {\bf 445}, 402 (2007).
\bibitem{Fol} S. F\"{o}llling, F. Gerbier, A. Widera, O. Mandel, T. Gericke, and I. Bloch, Nature {\bf 434}, 481 (2005).
\bibitem{Rom} T. Rom, Th. Best, D. Van Oosten, U. Schneider, S. F\"{o}llling, B. Paredes, and I. Bloch, Nature {\bf 444}, 733 (2006).

\end{thebibliography}
\end{document}